\documentclass[aps,twocolumn,pra,tightenlines,floatfix,showpacs]{revtex4}
\usepackage[dvips]{graphicx}
\usepackage[english]{babel}
\usepackage{amsmath}
\usepackage{amssymb}
\usepackage{times}
\usepackage{float}
\def\be{\begin{equation}}
\def\ee{\end{equation}}
\def\bea{\begin{eqnarray}}
\def\eea{\end{eqnarray}}

\def\d{\mbox{d}}
\newcommand{\ek}{\epsilon_{\mathbf{k}}}
\newcommand{\dbar}[1]{\bar{\bar{#1}}}

\newcommand{\mb}[1]{{\mathbf{#1}}}
\newcommand{\sumk}{{\sum_{\mathbf{k}}}}

\renewcommand{\text}[1]{{#1}}

\begin{document}

\title{Impurity effects on BCS-BEC crossover in ultracold atomic Fermi
  gases}

\author{Yanming Che}

\affiliation{Zhejiang Institute of Modern Physics and Department of Physics,
Zhejiang University, Hangzhou, Zhejiang 310027, China}
\affiliation{Synergetic Innovation Center of Quantum Information and
  Quantum Physics, Hefei, Anhui 230026, China}

\author{Qijin Chen}
\email[Corresponding author: ]{qchen@zju.edu.cn}
\affiliation{Zhejiang Institute of Modern Physics and Department of Physics,
Zhejiang University, Hangzhou, Zhejiang 310027, China}
\affiliation{Synergetic Innovation Center of Quantum Information and
  Quantum Physics, Hefei, Anhui 230026, China}

\date{\today}

\begin{abstract}
  We present a systematic investigation of the effects of
  ``nonmagnetic'' impurities on the $s$-wave BCS-BEC crossover in
  atomic Fermi gases within a pairing fluctuation theory. Both pairing
  and impurity scattering $T$-matrices are treated self-consistently
  at the same time. While the system is less sensitive to impurity
  scattering in the Born limit, for strong impurity scatterers, both
  the frequency and the gap function are highly renormalized, leading
  to significant suppression of the superfluid $T_c$, the order
  parameter and the superfluid density. We also find the formation of
  impurity bands and smearing of coherence peak in the fermion density
  of states, leading to a spectrum weight transfer and finite lifetime
  of Bogoliubov quasiparticles. In the BCS regime, the superfluidity
  may be readily destroyed by the impurity of high density. In
  comparison, the superfluidity in unitary and BEC regimes is
  relatively more robust.

\end{abstract}


\maketitle

\section{INTRODUCTION}
\label{sec:I}

Ultracold atomic Fermi gases have been a rapidly growing field over
the past decade, and have attracted enormous attentions from various
disciplines including condensed matter, atomic and molecular physics,
nuclear matter and astrophysics. Owing to the high tunability of
multiple parameters, atomic Fermi gases have become a prototype for
quantum simulations of a vast range of existing quantum systems in,
e.g., condensed matter and for engineering highly exotic quantum
states \cite{ChenReports}. One such system is superconductors with a
tunable interaction strength. Despite that atomic Fermi gases can be
prepared as a clean system without an impurity, impurities are hard to
avoid in a typical condensed matter system, including the most
important and widely studied high $T_c$ cuprate and Fe-based
superconductors. Therefore, study of the impurity effects on the
superfluidity and pairing phenomena using an atomic Fermi gas is very
important.

Associated naturally with a two-component Fermi gas is the physics of
superfluidity and pairing, whose counterpart condensed matter system
is superconductivity. The related impurity effects in superconductors
have also been an important subject, including superconducting
alloys\cite{AG-JETP}, disordered high $T_c$
superconductors\cite{Pan,Chen-PRB}, disordered superconducting thin
films and the disorder induced superconductor-insulator transition
\cite{LevIoffe-NaturePhys,Trivedi-NaturePhys,Dubi-Nature,Feigelman-PRL},
etc.  While the impurity effects in conventional phonon-mediated
$s$-wave superconductors have been understood fairly well, the
pseudogap phenomena in ($d$-wave) cuprate superconductors have
introduced further complexity \cite{HongDing-Nature}. Unlike a typical
superconductor, the pairing interaction strength in atomic Fermi gases
can be tuned via a Feshbach resonance from weak to strong, effecting a
crossover from BCS superfluidity to Bose-Einstein condensation (BEC).
Pseudogap phenomena have been widely recognized as the pairing
strength become strong. It is thus interesting to go beyond weak
coupling BCS theory and study the impurity effects in the presence of
strong pairing. Indeed, experimentally, impurities can be introduced
via doping atoms of foreign elements \cite{LiHan-NJP} or using a
random optical potential \cite{Laurent-naturephys}.

In a conventional $s$-wave BCS superconductor, weak impurities
renormalize the frequency $\omega$ and gap $\Delta$ in the exact same
fashion \cite{AG-book}, such that their effects are canceled out in
the gap equation, leading to the Anderson's theorem \cite{Anderson},
with an unchanged superconducting transition temperature $T_c$.  For a
$d$-wave superconductor, it has been known that impurities often lead
to a quadratic temperature dependence for the low temperature London
penetration depth or superfluid density. Chen and Schrieffer
\cite{Chen-PRB} have studied the effects of nonmagnetic impurities for
a $d$-wave superconductor on a quasi-two-dimensional lattice from the
Born to unitary limits of the impurity scattering strength, and for
both weak and strong pairing strengths. On the other hand, since no
strong pairing $s$-wave superconductors have been found thus far,
there has been very few studies of the impurity effects in strong
pairing $s$-wave superconductivity. Orso studied BCS-BEC crossover in
a random external potential \cite{PhysRevLett250402}, based on the
Nozieres and Schmitt-Rink (NSR) theory \cite{NSR}, which has been
known to suffer from inconsistencies between its $T_c$ equation and
fermion number equation in terms of the self-energy contributions of
pairing fluctuations. Han and Sa de Melo \cite{LiHan-NJP} studied the
BCS-BEC crossover in the presence of disorder, using functional
integrals and a local density approximation, which requires the
interaction range of the disorder be much larger than the pair size.
Recently, Strinati and coworkers \cite{Strinati} studied the impurity
effects in the context of BCS-BEC crossover, but at the lowest order,
using a diagrammatic approach, which does not have a pseudogap in the
$T_c$ equation even in the strong pairing regime, where the presence
of a pseudogap has been established experimentally. At the same time,
higher order contributions from impurity scattering in the strong
scattering regime are missing in their treatment, and the impurity and
pairing $T$-matrices are not treated in a self-consistent fashion,
either.

In this paper, we will present a systematic treatment of the impurity
effects on a two-component ultracold atomic Fermi gas as a function of
the impurity strength, impurity concentration, and pairing interaction
strength, in the case of $s$-wave pairing throughout the BCS-BEC
crossover. We will use the formalism developed in
Ref.~\cite{Chen-PRB}, where the pairing fluctuations and nonmagnetic
impurity $T$-matrix are treated self-consistently. While the original
formalism was applied to $d$-wave pairing on a quasi-two-dimensional
lattice, relevant to cuprate superconductors, here we apply it to
$s$-wave pairing throughout the entire BCS-BEC crossover in a three
dimensional (3D) atomic Fermi gas.  Unlike the nodal $p$-wave
\cite{Hirschfeld} and $d$-wave cases \cite{Chen-PRB}, where the
gap renormalization vanishes, for the $s$-wave pairing, both of the
frequency and gap renormalization induced by impurities are
present. Only for weak impurity scattering (i.e., the Born limit),
where the impurity potential may be treated at the Abrikosov-Gor'kov
(AG) level, the frequency and gap renormalization factors are exactly
the same so that the Anderson's theorem is valid
\cite{AG-book,Zhu-RMP}.

Our main results are as follows: (a) In the presence of strong
impurity scattering, the frequency and the gap function are highly
renormalized, leading to significant suppression of the superfluid
$T_c$.  (b) In the BCS regime, impurities induce impurity bands,
subgap states and strong smearing of the coherent peak (CP), thus the
superfluidity may be readily destroyed by the impurity. Besides, we
find an effective power law dependence of $T_c$ as a function of
pairing strength\cite{PRL-AL}.  (c) Superfluidity in the unitary and
BEC regimes is relatively more robust than in the BCS regime.  (d)
$S$-wave pairing is less sensitive to impurity than its $d$-wave
counterpart\cite{Chen-PRB}.  (e) Strong impurity scatterers are much
more effective than weak scatterers in the Born limit, in suppressing
$T_c$, order parameter, and the superfluid density. 

It should be noticed that there are also other theoretical approaches
toward the interplay of BCS-BEC crossover and impurity, mainly using
functional integral formalism and the replica trick
\cite{LiHan-NJP,PhysRevLett250402}.

The rest of this paper is arranged as follows. In Section \ref{sec:II}
we briefly capitulate the theoretical formalism developed in
Ref.~\cite{Chen-PRB}, with a focus on the main results and the
differences between the $s$-wave atomic Fermi gases and the $d$-wave
cuprate superconductors.  In Section \ref{sec:III} we numerically
solve the set of equations to get various impurity renormalization
effects on density of states (DOS), $T_c$, gaps, and superfluid
density, etc, throughout the BCS-BEC crossover. Finally we discuss the
results and experiment related issues.

\section{Theoretical FORMALISM}
\label{sec:II}
\subsection{Frequency and gap renormalizations}

The formalism for BCS-BEC crossover at finite temperature in a clean
system can be found in Section \ref{sec:II}(A) of
Ref.~~\cite{Chen-PRB}.  Here for atomic Fermi gases of $^6$Li or
$^{40}$K, we take the free fermion dispersion
$\epsilon^0_{\mathbf{k}}=\mathbf{k}^2/(2m)$, and a contact potential for
the $s$-wave pairing interaction $V_\mathbf{k,k'}^{} =
g\varphi_\mathbf{k}^{}\varphi_\mathbf{k'}^{}$, with
$\varphi_{\mathbf{k}} = 1$, where $m$ is the atomic mass and we take
$\hbar=k_B^{}=1$ as usual. The ultraviolet divergence in the gap
equation, caused by the unphysical contact potential, can be
regularized in a standard way so as to replace $g$ with $1/k_F^{}a$
using the Lippmann-Schwinger equation \cite{ChenReports}
\begin{equation}
\frac{m}{4\pi a} = \frac{1}{g} + \sum_{\mathbf{k}} \frac{1}{2\epsilon^{0}_{\mathbf{k}}},
\label{eq:LS}
\end{equation}
where $k_F^{}$ is the Fermi wave vector and $a$ is the two-body $s$-wave
scattering length. Now by solving self-consistently the gap equation,
atomic number equation and pseudogap equation, one can study BCS-BEC
crossover at finite temperature in atomic Fermi gases as a function of
$1/k_F^{}a$.

The impurity Hamiltonian is given by
\begin{equation}
 H_I = \sum_i\int \d\mathbf{x}\;
\psi^\dagger(\textbf{x})u(\textbf{x}-\textbf{x}_i)\psi(\textbf{x}) \:,
\end{equation}
with $u(\textbf x-\textbf{x}_i)=u\delta(\textbf x-\textbf{x}_i)$,
where $\textbf{x}_i$ denotes independent, randomly distributed impurity
sites. We refer to these impurities as ``nonmagnetic'' in the sense
they cannot convert one species of atoms into the other, similar to a
superconductor where a nonmagnetic impurity does not cause spin
flips.  

At the AG level \cite{AG-book, AG-JETP}, impurities in a $s$-wave BCS
superconductor only induce frequency and gap function renormalization,
leading to the Anderson's theorem for weak impurities. In
Ref.~\cite{Chen-PRB}, Chen and Schrieffer went beyond the AG level,
and considered impurities of arbitrary strength and variable pairing
interactions by treating the impurity $T$-matrix and pairing
fluctuations self-consistently at the same time. Now we shall present
the main results of the formalism, while detailed derivations can be
found in Ref.~\cite{Chen-PRB}.

The frequency and gap renormalizations now are given in terms of the
impurity $T$-matrices, $T_{\omega}$ and $T_{\Delta}$ (and its complex
conjugate), by
\begin{subequations}
\begin{equation}
i\tilde{\omega} = i\omega - \Sigma_\omega \,, \quad
i\tilde{\underline{\omega}} = -i\omega - \Sigma_{-\omega} \,,
\end{equation}
\begin{equation}
\tilde{\Delta}_\mb{k} = \Delta_\mb{k} + \Sigma_\Delta \,,\quad
\tilde{\Delta}_\mb{k}^* = \Delta_\mb{k}^* + \Sigma_{\Delta^\dag} 
\end{equation}
\end{subequations}
where $\Delta_\mb{k}^{} = \Delta\varphi_\mathbf{k}^{}$, $\Sigma_\omega
= n_i T_\omega$ and $\Sigma_\Delta = n_i \Delta T_\Delta$, with $n_i$
being impurity density. Here $\tilde{\underline{\omega}}=
\widetilde{(-\omega)}$.  Now except that $i\tilde{\omega}$ and
$\tilde{\Delta}_\textbf{k}$ acquire new expressions, the Green's
function $G$, Gor'kov function $F$, and the pair susceptibility
$\chi(Q)$ remain formally the same in terms of $i\tilde{\omega}$ and
$\tilde{\Delta}_\textbf{k}$.  These expressions reduce to the AG-level
results in the lowest order (Born limit).

It should be pointed out \cite{Chen-PRB} that here $\Delta$ is the
excitation gap, related to the order parameter $\Delta_{sc}$ and
pseudogap $\Delta_{pg}$ via $\Delta^2_{} = \Delta_{sc}^2 +
\Delta_{pg}^2$.

\begin{figure}
\centerline{\includegraphics[width=3.2in]{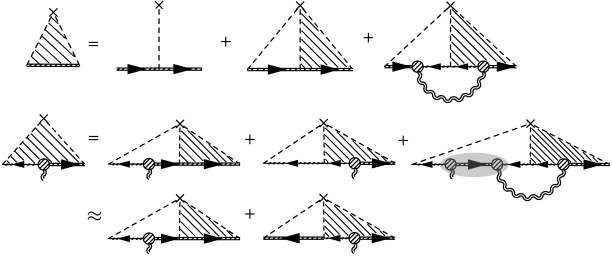}}
\caption{Feynman diagrams for the regular impurity $T$-matrix
  $T_\omega$ and the anomalous impurity $T$-matrix
  $T_{\Delta^\dagger}$. The crosses denote the impurities and the
  dashed lines represent the impurity potential.  The dressed thin
  solid, thick solid lines represent impurity dressed noninteracting
  and interacting fermion propagators, respectively.  The shaded
  elliptical region denotes self-consistent impurity dressing of the
  double pairing vertex structure. See Ref.~\cite{Chen-PRB} for
  details.}
\label{fig:Impurity_Tmatrices}
\end{figure}

Shown in Fig.~\ref{fig:Impurity_Tmatrices} are the Feynman diagrams for
the impurity $T$-matrices $T_{\omega}$ and $T_{\Delta^\dag}$,
respectively, where $T_{\Delta^\dag}$ is the complex conjugate
of $T_{\Delta}$.
Note that for the impurity potential we use here, with $u(\mb k, \mb
k^{\prime})=u$, the impurity $T$-matrices only acquire a dynamical
structure and are independent of the fermion momentum. Thus,
$T_\omega$ and $T_{\Delta}$ can be decoupled as
\begin{subequations}
\label{eq:ImpT}
\begin{equation}
  T_\omega = \frac{u (1-u\dbar{G}_{-\omega})}{\left( 1-u
      \dbar{G}_\omega\right) \left( 1-u
      \dbar{G}_{-\omega}\right) + u^2 \dbar{F}_\omega
    \dbar{F}^\dagger_\omega} \,,
\label{eq:Tw}
\end{equation}
and
\begin{eqnarray}
  T_{\Delta^\dag} (\omega-\Omega, \omega|Q) &=& \frac{u^2 \sum_{\bf
      k} G_0(Q-K) \Gamma_0^{} (K|Q) G(K)} {\left( 1-u
      \dbar{G}_\omega\right) \left( 1-u
      \dbar{G}_{-\omega}\right) + u^2 \dbar{F}_\omega
    \dbar{F}^\dagger_\omega }\nonumber \\ &&{}\times \frac{ 1-u
    \dbar{G}_{-\omega} }{  1-u \sumk G(Q-K) } \,,
\label{eq:TDeltaQ}
\end{eqnarray}
where we have used a four-momentum notation, $K\equiv ({\bf k},
i\omega_l), Q\equiv ({\bf q}, i\Omega_n)$, with $\omega_l$ and
$\Omega_n$ being the odd and even Matsubara frequencies, respectively.
Here we have also defined the impurity averaged Green's functions
$\bar{\bar{G}}_\omega = \sum_{\mb{k}}G(K)$ and
$\bar{\bar{G}}_{-\omega} = \sum_{\mb{k}}G(-K)$, and the anomalous 
Green's function $\dbar{F_\omega}$ (and its complex conjugate
$\dbar{F_\omega}^\dagger$) as
\begin{equation}
 \dbar{F_\omega}^\dag = \sum_{\mathbf{k}} F^\dag(K) = \sum_{\mathbf{k}} \Delta
\Gamma_0^{}(K) G_0(-K)G(K) \,,
\label{eq:Fw}
\end{equation}
where $\Gamma_0(K)=\tilde{\Delta}_\mb{k} / \Delta$ is the renormalized
pairing vertex function. For isotropic impurity scattering,
$T_\omega^{}$ and $T_{\Delta^\dag}$ are independent of momentum.

In the static limit, $Q\rightarrow 0$, the
expression for $T_{\Delta^\dag}$ becomes
\begin{equation}
    T_{\Delta^\dag}(\omega) = \frac{u^2 \dbar{F}_\omega / \Delta} {\left( 1-u
    \dbar{G}_\omega\right) \left( 1-u
    \dbar{G}_{-\omega}\right) + u^2 \dbar{F}_\omega
    \dbar{F}^\dagger_\omega } \,.
\label{eq:TDelta}
\end{equation}
\end{subequations}
\emph{Unlike the cases of $d$-wave \cite{Chen-PRB} and $p$-wave pairing}
\cite{Hirschfeld}, where $\dbar{F}_\omega \equiv 0$ so that this gap
renormalization vanishes, here for our $s$-wave pairing, $\varphi_{\mb
  k}=1$, the gap renormalization is given by
\begin{equation}
\tilde{\Delta}=\Delta+\Sigma_\Delta = \Delta \Gamma_0 (\omega) \, ,
\end{equation}
where
\begin{equation}
\Gamma_0 (\omega)=1+ n_i T_\Delta (\omega) \,.
\label{eq:Gammaw}
\end{equation}
For the momentum independent vertex function in Eq.~(\ref{eq:Gammaw}),
using the expression for $T_\Delta$ in Eq.~(\ref{eq:TDelta}) and
the impurity averaged Gor'kov function in Eq.~(\ref{eq:Fw}),
$\Gamma_0 (\omega)$ can be written explicitly as
\begin{equation}
\Gamma_0 (\omega) = \frac{1}{1-n_i u^2 \lambda(\omega)},
\label{eq:gamma_w}
\end{equation}
where
\begin{equation}
\lambda(\omega) = \frac{\sum_{\mb{k}} G (-K) G^{}_0 (K)}{{\cal D}_\omega}\,,
\label{eq:lambda}
\end{equation}
%
with
\begin{equation}
  {\cal D}_\omega = \left( 1-u
    \dbar{G}_\omega\right) \left( 1-u
    \dbar{G}_{-\omega}\right) + u^2 \dbar{F}_\omega
    \dbar{F}_\omega^\dagger \, .
\label{eq:D}
\end{equation}
Finally, the full Green's function is given by
\begin{equation}
G(K) = \frac{i\tilde{\underline{\omega}} -\ek}
{(i\tilde{\omega}-\ek) (i\tilde{\underline{\omega}} -\ek)  +
  \tilde{\Delta}_\mb{k}^* \tilde{\Delta}_\mb{k}} \,.
\label{eq:GF0_AC}
\end{equation}

\subsection{Analytical continuation and spectral representation}
\label{sec:IIB}

In order to numerically calculate the impurity renormalization
functions, the Matsubara frequencies need to be analytically continued
to the real frequencies, $i\omega_l \longrightarrow \omega + i 0^+$. 
In general, we have $i\tilde{\underline{\omega}} \neq
-i\tilde{\omega}$, due to the absence of the particle-hole
symmetry. Therefore, both the positive and negative frequencies should
be analytically continued at the same time.  For $l> 0$,
$i\tilde{\omega}_l \rightarrow
\omega_+^R = \omega_+ + i\Sigma''_+ $, and
$i\tilde{\underline{\omega}}_l \rightarrow
\omega_-^A = \omega_- - i \Sigma''_- $.  For $l^\prime = -l < 0$,
$i\tilde{\omega}_{l^\prime} \rightarrow
\omega_-^R = \omega_- + i \Sigma''_-$ and
$i\tilde{\underline{\omega}}_{l^\prime} \rightarrow
\omega_+^A = \omega_+ - i\Sigma''_+$. Here $\omega_\pm = \pm \omega -
\Sigma^\prime_\pm$, and we choose $\omega> 0$ and $\Sigma_\pm''>0$.
The superscripts $R$ and $A$ denote retarded and advanced analytical
continuations, respectively. We obtain
\begin{eqnarray}
\dbar{G}^R_{\omega> 0}\!\! &=&\!\! \sumk \frac{\omega_- - i\Sigma_-'' -\ek}
{(\omega_+\! +\! i \Sigma_+''\! -\ek) (\omega_-\! -\! i\Sigma_-''\! -\ek)\! +\!
  \Delta^2|\Gamma_0^R (\omega)|^2}\,, \nonumber\\
\dbar{G}^R_{-\omega< 0}\!\! &=&\!\! \sumk \frac{\omega_+ - i\Sigma_+'' -\ek}
{(\omega_-\! +\! i \Sigma_-''\! -\ek) (\omega_+\! -\! i\Sigma_+''\! -\ek)\! +\!
  \Delta^2|\Gamma_0^R (\omega)|^2}\,, \nonumber\\
\Sigma^R_{\omega>0}\!\! &=& \!\!\frac{n_i u(1 - u \dbar{G}^R_{-\omega})}{{\cal D}^R_\omega} =
\Sigma_+^\prime - i \Sigma_+''\,, \nonumber\\
\Sigma^R_{-\omega<0}\!\! &=& \!\!\frac{n_i u (1 - u \dbar{G}^R_{\omega})}{{\cal D}^R_\omega} =
\Sigma_-^\prime - i \Sigma_-'' \,,
\label{eq:renorm}
\end{eqnarray}
where $\epsilon_{\textbf k}=\textbf{k}^2/(2m)-\mu$ with $\mu$ being
the chemical potential, and $ |\Gamma_0^R (\omega)|^2 = \Gamma_0^R
(\omega) \Gamma^{A}_0 (\omega) = \Gamma^{\prime 2}_0 (\omega) +
\Gamma''^2_0 (\omega) $. From Eq.~(\ref{eq:gamma_w}), we have 
\begin{equation}
\Gamma_0^R (\omega) = \Gamma^{\prime}_0 (\omega) +
i\Gamma''_0 (\omega) = \frac{1}{1-n_i u^2 \lambda^R(\omega)},
\label{eq:gamma0}
\end{equation}
where $\lambda^R(\omega)$ can be calculated from
Eqs.~(\ref{eq:lambda})-(\ref{eq:GF0_AC}). Note that here the gap
renormalization function $\Gamma_0 (\omega)$ involves pairing between
four-momenta $\pm K$. Therefore, we have the symmetry $\Gamma^*_0
(\omega)=\Gamma_0 (-\omega)$. This is different from the frequency
renormalization $\Sigma_{\omega}$.

Equations (\ref{eq:renorm}) and (\ref{eq:gamma0}) form a closed set
for solving for the six variables $\Sigma'_ {\pm \omega},
\Sigma''_{\pm \omega}, \Gamma'_0 (\omega),\Gamma''_0 (\omega)$ as a
function of $\omega$. In comparison with the $d$-wave case in
Ref.~\cite{Chen-PRB}, here we have two more extra equations to solve.

For the 3D Fermi gas and the contact impurity potential we consider
here, the real part of $\dbar{G}^R_{\omega}$ diverges, caused by the
momentum integral over $\textbf{k}$ far away from the Fermi surface.
In the AG theory, this ultraviolet divergence is absorbed into the
chemical potential $\mu$, signifying an additive correction, $\delta
\mu$, to the chemical potential \cite{AG-book}. Here we adopt a
similar regularization scheme, and replace $\dbar{G}^R_{\omega}$ with
\begin{equation}
\dbar{{\cal G}}^R_{\omega}=\dbar{G}^R_{\omega}-\dbar{G}^R_{\omega=0}\,.
\label{eq:regularization}
\end{equation}
It is easy to show that $\dbar{G}^R_{\omega=0}$ is real and thus can
be fully absorbed into a renormalized chemical potential. In the
lowest order (Born limit), ${\cal D}_\omega \approx 1$ and
$\Sigma_\omega \sim n_i u^2 \dbar{G}^R_{-\omega}$, so that our
regularization scheme reduces to that of the AG theory, with $\delta
\mu \sim n_i u^2 \dbar{G}^R_{\omega=0}$. Beyond the Born limit, the
corrections to $\Sigma_{\omega}$ and $\Gamma_0 (\omega)$ caused by the
regularization are proportional to $1/\dbar{G}^R_{\omega=0}$ and
$1/(\dbar{G}^R_{\omega=0})^2$, respectively, which are
negligible.

With the renormalization functions $\Sigma_\omega$ and
$\Gamma_0 (\omega)$, one can calculate the pair susceptibility
\begin{equation}
\chi (Q) =  \sum_K \Gamma_0^{}(K|Q)G(K)G_0(Q-K),
\end{equation}
whose real and imaginary parts are given respectively by
\begin{subequations}
\begin{eqnarray}
\lefteqn{\chi^\prime(\Omega+i 0^+, \mb{q})}\nonumber\\
&=& \mbox{Im} \sumk \int_{-\infty}^\infty \frac{d\omega}{2\pi} \Big\{
  G^R(\omega,\mb{k}) G^R_0(\Omega-\omega, \mb{q-k})\nonumber\\&&{}
\times \left[ f(\omega-\Omega)-f(\omega) \right] +  G^R(\omega,\mb{k})
G^A_0(\Omega-\omega, \mb{q-k}) \nonumber\\&&{} \times
  \left[1-f(\omega) - f(\omega-\Omega)\right]\Big\} \Gamma^R_0 (\omega|Q) \,,
\label{eq:RechiQ}
\end{eqnarray}
and
\begin{eqnarray}
\lefteqn{\chi''(\Omega+i 0^+, \mb{q})}\nonumber\\
& =& - \mathrm{Im}\, \sumk \int_{-\infty}^\infty \frac{d\omega}{2\pi}
 G^R(\omega,\mb{k}) A_0(\Omega-\omega, \mb{q-k}) 
\nonumber\\&&{} \times \left[
    f(\omega-\Omega)-f(\omega) \right] \Gamma_0^R (\omega|Q)\,,
\label{eq:ImchiQ}
\end{eqnarray}
\end{subequations}
where $f(\omega)$ is the Fermi distribution function. Here $\Gamma_0^R
(\omega|Q)$ can be obtained from Eq.~(\ref{eq:TDeltaQ}) after
analytical continuation, with $\Gamma_0^R
(\omega|0)=\Gamma_0^R(\omega)$. And $A_0(\omega, \mb{k}) =
-2\,\mbox{Im}\, G^R_0(\omega, \mb{k})$ is the ``bare'' spectral
function.

For $Q=0$, we obtain
\begin{equation}
\chi(0)  = \mbox{Im} \sumk \int_0^\infty \frac{d\omega}{\pi}
  \frac{[1-2f(\omega)]\: \Gamma_0^R (\omega)}
  { C(\omega, \mb{k})}  \,,
\label{eq:chi-int}
\end{equation}
where $C(\omega, \mb{k})=(\omega_+ + i\Sigma_+''-\ek) (\omega_- -
i\Sigma_-'' - \ek) + \Delta^2|\Gamma_0^R (\omega)|^2$.

Substituting into the Thouless criterion, $1+g\chi(0)=0$, we have the
gap equation
\begin{equation}
-\frac{m}{4\pi a} = \sum_{\mathbf{k}} \left[\mbox{Im} \int_{-\infty}^\infty
\frac{d\omega}{2\pi}  \frac{\left[1-2f(\omega) \right] \Gamma_0^R (\omega) }
{C(\omega,\mb k)} - \frac{1}{2\epsilon^0_{\mathbf{k}}} \right]\,,
\label{eq:gapeq-imp}
\end{equation}
where we have used the Lippmann-Schwinger equation (\ref{eq:LS})
to replace $g$ with scattering length $a$.

Now the fermion number equation becomes
\begin{eqnarray}
n
&=& 2\sumk \int_{-\infty}^\infty \frac{d\omega}{2\pi}
A(\mb{k},\omega)f(\omega) \nonumber \\
&=& \int_{-\infty}^\infty \frac{d\omega}{\pi} N(\omega) f(\omega)\,,
\label{eq:number-imp}
\end{eqnarray}
where $A(\mb{k},\omega)= -2\, \mathrm{Im}\, G^R (\omega, \mb{k})$ is the
renormalized spectral function and
\begin{eqnarray}
  N (\omega) = \sumk A(\mb{k},\omega) =   -2\, \mathrm{Im}\, \dbar{G}^R_\omega
\label{eq:DOS}
\end{eqnarray}
is the density of states.  Next we evaluate the pseudogap, which
is given by
\begin{eqnarray}
\Delta^2_{pg} = -\sum_{\mb q} \int_{-\infty}^\infty \frac{d\Omega}{\pi}\,
 \mathrm{Im}\, t^R (Q) \,  b(\Omega) \, ,
\end{eqnarray}
where $b(x)$ is the Bose distribution function, and the retarded $T$
matrix $t^R (Q) = \left[ \chi(\Omega+ i 0^+, \mb{q})-\chi(0, \mb{0})
\right]^{-1}$.
In actual numerics, we follow Ref.~\cite{Chen-PRB} and Taylor expand
the inverse $T$ matrix, $t^{-1} (\Omega+i 0^+, \mb{q})$, which greatly
facilitates the computation.

\section{NUMERICAL RESULTS}
\label{sec:III}

The numerics is done as follows. First, for given (initial) values of
the parameters $\left[ \mu, T_c, \Delta \right]$, the renormalization
spectrum $\Sigma_\omega$ and $\Gamma_0 (\omega)$ are solved. Next,
these renormalization functions are substituted into the gap,
pseudogap and fermion number equations and an equation solver is used
to obtain $\mu(T_c)$, $T_c$, and $\Delta(T_c)$ at $T_c$ and gap, order
parameter as well as $\mu$ below $T_c$. With these newly obtained
parameters, the equation solver will repeat the above process, until
self-consistent solutions are obtained.

\subsection{Impurity renormalization functions and the density of states }
\label{subsec:DOS}

In this subsection, we numerically solve the the coupled equations for
$\left[\Sigma'_{\pm },\Sigma''_{\pm },\Gamma'_0, \Gamma''_0 \right]$
as a function of $\omega$ for given impurity levels and study the
impurity renormalization effects on the frequency and gap function, as
well as the DOS.

\begin{figure}
\centerline{\includegraphics[width=3.3in,clip]{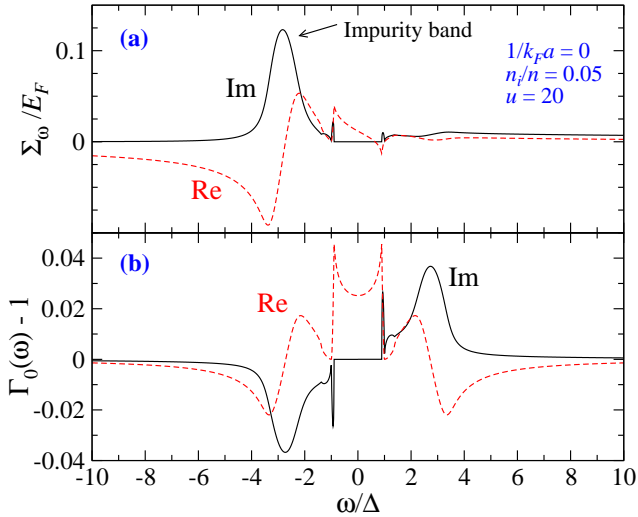}}
\caption{The impurity renormalization functions $\Sigma_\omega$ and
  $\Gamma_0 (\omega)$, in a unitary Fermi gas at an intermediate
  impurity scattering level, with $n_i=0.05n$, $u/E_F^{}=20$. Re (Im)
  denotes the real (imaginary) part of the functions. There are sharp
  features near $\omega =\pm \Delta$. The broad peak of $
  \Sigma^{\prime\prime}_\omega$ at $\omega/\Delta \approx -3$ (black
  curve) indicates the formation of an impurity band. The horizontal
  axes are rescaled by the gap $\Delta$ of a clean system.  }
\label{fig:imp_renorm}
\vspace*{0.1cm}
\end{figure}

We first present in Fig.~\ref{fig:imp_renorm} typical (a) frequency
and (b) gap (pairing vertex) renormalization functions $\Sigma_\omega$
and $\Gamma_0 (\omega)$, respectively. Shown here are the functions
for a unitary Fermi gas at an intermediate impurity level $n_i/n =
5\%$ and impurity scattering strength $u/E_F^{}=20$, which is close to
the unitary scattering limit. The real (Re) and imaginary (Im) parts
of $\Sigma_\omega$ and $\Gamma_0 (\omega)$ are solved at the same
time, as explained above. The frequency axis is plotted in units of
the clean gap $\Delta=0.64E_F^{}$. From Fig.~\ref{fig:imp_renorm}(a),
one can easily spot an impurity band (IB) near $\omega \simeq
-3\Delta$, which manifests as a broad peak in
$\Sigma^{\prime\prime}(\omega)$. The peak location is given by the
zero point of Re$\,{\cal{D}}_{\omega}$, leading to a peak in
$\Gamma^{\prime\prime}_0 (\omega)$ as well, which shares the same
denominator. Interestingly, even for positive $u$, the IB may occur on
the negative $\omega$ side as well, due to the presence of a nonzero
$\dbar{F}_\omega$ in ${\cal D}_{\omega}$ and particle-hole
mixing. Indeed, as one can easily see from Eq.~(\ref{eq:Tw}), for
large $|u|$, the sign of $u$ becomes almost irrelevant. This should be
contrasted to the $d$-wave case in Ref.~\cite{Chen-PRB}. There are
sharp features related to the pairing gap edge at $\omega=\pm \Delta$,
esp. in the pairing vertex renormalization function
$\Gamma_0(\omega)$. With the impurity configuration in
Fig.~\ref{fig:imp_renorm}, inside the gap (i.e., $|\omega| < \Delta$),
the imaginary parts of both $\Sigma_\omega$ and $\Gamma_0 (\omega)$
are essentially zero. In addition, as expected, the impurity
renormalization effects are mostly in the low frequency regions, and
decreases with a power law of $|\omega|$ at sufficient high
frequencies.  Note that for clarity, here we plot $\Gamma_0
(\omega)-1$ in Fig.~\ref{fig:imp_renorm}(b), as the renormalization is
small in comparison to its unrenormalized value, $\Gamma_0
(\omega)=1$. As a consistency check, we note that
Fig.~\ref{fig:imp_renorm}(b) obeys the symmetry
$\Gamma_0(-\omega)=\Gamma_0^*(\omega)$.

As can be seen from Fig.~\ref{fig:imp_renorm}, the real part of the
impurity scattering is in general small compared to the unrenormalized
part. For the frequency, $\Sigma'_\pm$ constitutes only a small
perturbation to $\omega$, so that the main impurity effect resides in
the imaginary parts. Inside the main band ($\omega >
-\sqrt{\mu^2+\Delta^2}$), a large $\Sigma''_\pm$ means a large
spectral weight loss, whereas outside the main band, it means a large
spectral weight gain. The imaginary parts increase with the impurity
density $n_i$.

For weak scattering in the Born limit, the impurity band does not
exist. Only when the scattering strength $|u|$ becomes large enough
are there significant spectral weight gain outside the main band.
The location of the impurity band (if it exists) is largely determined
by the impurity strength $u$, whereas the impurity density $n_i$
affects the magnitude of $\Sigma^{\prime\prime}(\omega)$ and
$\Gamma^{\prime\prime}_0 (\omega)$ and the spectral weight of the
impurity band. The impurity band becomes prominent only when it is
located outside the main band. In the BCS regime, the gap $\Delta$
becomes small. Once the impurity band is clearly visible, it will
appear on the far left side in a plot such as
Fig.~\ref{fig:imp_renorm}.

\begin{figure}
\centerline{\includegraphics[width=3.3in,clip]{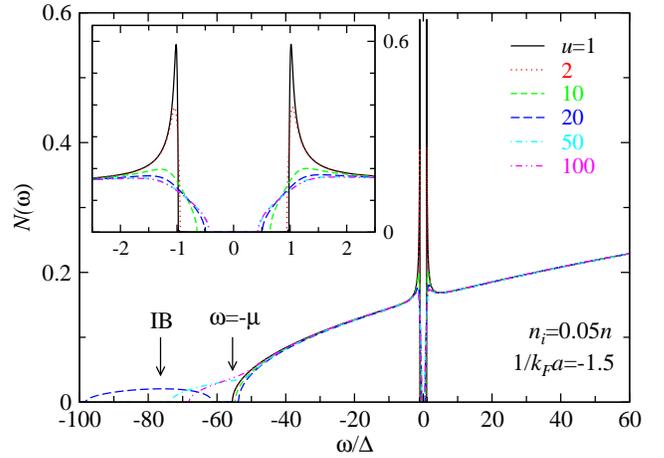}}
\caption{Effects of different impurity scattering strength $u$ from
  the Born limit $u/E_F^{}=1$ to the unitary limit $u/E_F^{}=100$ on
  the fermion DOS $N(\omega)$ in the BCS regime ($1/k_F^{}a=-1.5$),
  with impurity density $n_i^{}=0.05n$.  The impurity band (IB) splits
  from the main band (blue curve) for an intermediate value of
  $u/E_F^{}=20$.  The inset presents the details of the coherence
  peaks, sharing the same axis labels. Here $\Delta \approx 0.018
  E_F^{}$. }
\label{fig:dos_BCS1}
\vspace*{0.3cm}
\end{figure}

Next we show in Fig.~\ref{fig:dos_BCS1} the effects of an increasing
impurity scattering strength $u$ on the DOS $N(\omega)$ for
$1/k_F^{}a=-1.5$ in the BCS regime, where the gaps are relatively small,
from the Born limit $u/E_F^{}=1$ to the unitary scattering limit
$u/E_F^{}=100$.  Here we choose a representative, intermediate impurity
density, $n_i^{}=0.05n$.  We show details of the coherence peak in the
inset. The location $\omega=-\mu$ indicates roughly the bottom of the
main band. It is evident that weak scatterers are not effective in
destroying the coherence peaks.  Indeed, this is in agreement with the
Anderson's theorem for weak impurities based on the AG level
treatment. However, when the impurity strength $u$ increases, say,
beyond 10, the coherence peaks become smeared out quickly, and
significant spectral weight is now moved inside the gap.  For
sufficiently strong $u$ and high density $n_i^{}$, the gap will be filled
up so that the superfluidity is destroyed. This should also be
compared with the $d$-wave case, where impurities in the Born limit
are found to be effective in smearing out the coherence peaks
\cite{Chen-PRB}.  Note that as mentioned earlier, the sign of $u$ is
nearly irrelevant for our short-range $s$-wave pairing. Therefore, we
plot here only curves for positive $u$.

One prominent feature in Fig.~\ref{fig:dos_BCS1} is the presence of
the impurity band for large $u$. More interestingly, the location of
the band does not move monotonically to the negative frequencies with
increasing $u$. In Fig.~\ref{fig:dos_BCS1}, the IB is well split from
the main band for $u=20$ (blue dashed curve), but partially overlaps
with the main band for the larger $u=50$ (cyan dot-dashed) and $u=100$
(magenta double-dot-dashed curve). Such a nonmonotonic behavior was
not seen for the $d$-wave case \cite{Chen-PRB}.

\begin{figure}
\centerline{\includegraphics[width=3.3in,clip]{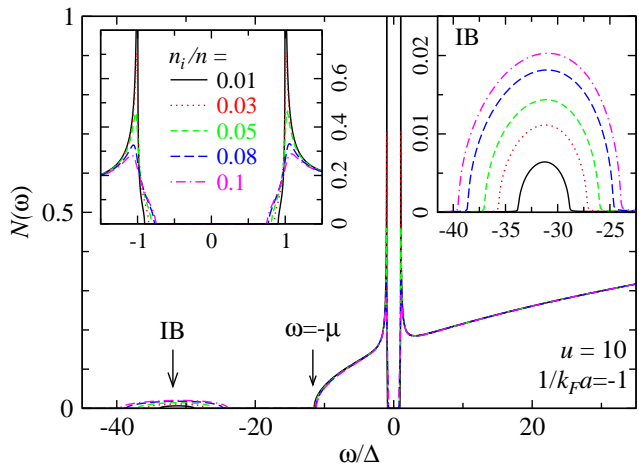}}
\caption{Effects of impurity scattering on the fermion DOS in the BCS
  regime ($1/k_F^{}a=-1$), with impurity scattering strength $u=10E_F^{}$
  and impurity density varying from $n_i^{}=0.01n$ to
  $n_i^{}=0.1n$. Detailed structure of $N(\omega)$ near the gap edge, and
  the impurity band are shown in the left and right insets,
  respectively. Here $\Delta \approx 0.085 E_F^{}$. }
\label{fig:dos_BCS2}
\vspace*{0.2cm}
\end{figure}

Shown in Fig.~\ref{fig:dos_BCS2} are the effects of increasing
impurity density $n_i^{}/n$ from $0.01$ to $0.1$ on the DOS
$N(\omega)$ for $1/k_F^{}a=-1$ in the BCS regime, with a fixed
$u=10E_F^{}$.  Shown in the left and right insets are the magnified
view of $N(\omega)$ for the coherence peaks and the impurity band
below the main band. Here the IB is well separated from the main band,
with spectral weight given by $2n_i^{}$.  The increasing impurity
density also serves to smear out and suppress the coherence peaks.
For sufficient high impurity $n_i^{}$, the superfluidity will be
destroyed. When comparing the coherence peaks for the $n_i^{}/n=0.05$
and $u/E_F^{}=10$ case between Figs.~\ref{fig:dos_BCS1} and
\ref{fig:dos_BCS2}, it is easy to conclude that a larger gap is more
robust against impurity scattering.

\begin{figure}
\centerline{\includegraphics[width=3.3in,clip]{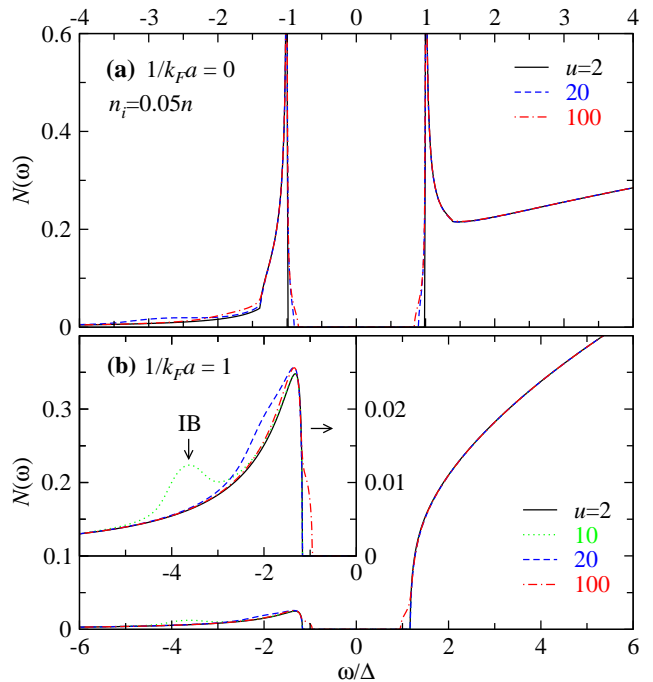}}
\caption{Effects of impurity scattering on the fermion DOS at (a)
  $1/k_F^{}a=0$ and (b) 1, corresponding to unitary and BEC cases,
  respectively, with the impurity density $n_i^{}=0.05n$, for
  different impurity scattering strength $u$ varying from the Born
  limit $u/E_F^{}=2$ to the unitary limit $u/E_F^{}=100$. Here
  $(\mu/E_F^{}, \Delta/E_F^{}) =(0.62,0.64)$ and (-0.80, 1.33), and
  the main band bottom is located at $\omega/\Delta\approx -1.4$ and
  1.17 for (a) and (b), respectively.}
\label{fig:dos_unitary_bec}
\vspace*{0.3cm}
\end{figure}

In Fig.~\ref{fig:dos_unitary_bec} we show the effects of varying
impurity scattering strength $u$ on the DOS in the (a) unitary and (b)
BEC regimes, respectively, with impurity density $n_i^{}/n=0.05$. With
substantially larger gaps, the DOS in these two regimes are very
robust against impurity effects.  Indeed, only minor smearing of
coherence peak can be found in the unitary case in
Fig.~\ref{fig:dos_unitary_bec}(a). For the BEC case with $1/k_F^{}a=1$
in Fig.~\ref{fig:dos_unitary_bec}(b), the chemical potential
$\mu/E_F^{}=-0.80$ is negative, so that there exists no underlying
Fermi surface. As a result, there are no coherence peaks in the clean
limit. The spectral weight below the bottom of the main band is mainly
a result of particle-hole mixing in both cases, with a power law tail
$N(\omega) \propto |\omega|^{-3/2}$ towards $\omega \rightarrow
-\infty$. Nevertheless, signatures of impurity band on top of this
power law tail can be seen for $u/E_F^{}=10$ (green dotted) and 20
(blue dashed curves). In addition, it is clear that a larger $u$ is
more effective in moving the spectral weight to within the gaps.
Though the DOS in these two regimes is not as sensitive to impurities
as in BCS regime, the finite $\Sigma(\omega)$ and $\Gamma_0^{}
(\omega)$ as well as the finite fermion pair lifetime (caused by
impurities) may also affect the superfluid $T_c^{}$ and other
superfluid properties.

\subsection{Effects of impurities on the behavior of $T_c$ in $s$-wave
  BCS-BEC crossover}
\label{subsec:Tc}

\begin{figure}
\centerline{\includegraphics[width=3.3in,clip]{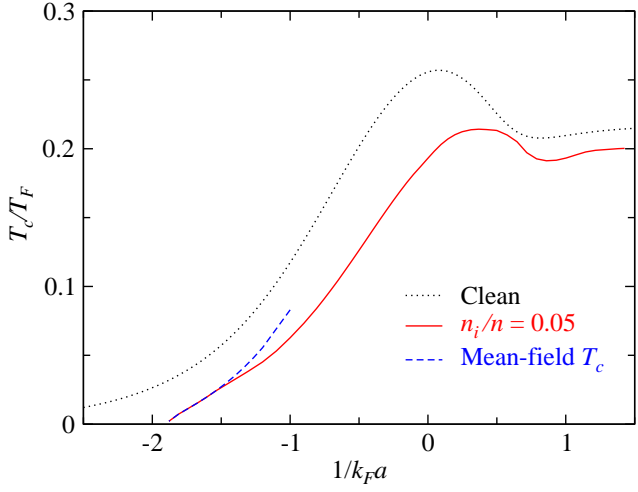}}
\caption{Behavior of $T_c$ throughout the BCS-BEC crossover in the
  clean limit (black dotted line) and in the presence of strong
  impurity scattering with density $n_i=0.05n$ and strength $u=100E_F^{}$
  (red solid line).  For comparison, also plotted is the mean field
  value of $T_c$ with the impurities (blue dashed line).}
\label{fig:CrossoverTc}
\vspace*{0.18cm}
\end{figure}

In this subsection we present the effects of impurities on $T_c$
throughout BCS-BEC crossover. Plotted in Fig.~\ref{fig:CrossoverTc} is
the behavior of $T_c$ as a function of $1/k_F^{}a$ from the BCS
through BEC regimes. For clarity, here we show only one case with a
representative impurity density $n_i^{}=0.05n$ in the unitary
scattering regime, $u/E_F^{}=100$ (red solid curve). For comparison,
we also show the $T_c$ curve in the clean system (black dotted line)
as well as the mean field result (blue dashed line). As one can expect
from previous figures, $T_c$ is suppressed by impurity scattering.
Furthermore, the relative suppression is much stronger in the BCS
regime than in the unitary and BEC regimes. In the deep BCS regime,
$T_c$ is suppressed down to zero by strong impurities, leading to an
effective power law dependence of $T_c$ on $1/k_F^{}a$ in the BCS
regime \cite{PRL-AL}.  In the pseudogap or crossover regime, the
maximum $T_c$ now shifts to the BEC side of the Feshbach resonance
(where $1/k_F^{}a=0$). This result is somewhat similar to the shift of
the $T_c$ curve by particle-hole fluctuations \cite{ParticleHole},
suggesting that even ``nonmagnetic'' impurities may to certain extent
have a pair-breaking effect. On the other hand, impurity scattering
and particle-hole fluctuations are very different. In the BCS regime,
while the latter simply reduces $T_c$ by a factor of 0.45, here strong
impurities in the unitary regime can destroy superfluidity completely
whereas weak impurities in the Born limit (not shown) may leave $T_c$
intact.  The result shown in Fig.~\ref{fig:CrossoverTc} should be
contrasted with the $d$-wave case \cite{Chen-PRB}. Due to the sign
change of the order parameter across the nodes in the momentum space,
impurity scattering is much more effective in destroying $T_c$
throughout the entire BCS-BEC crossover. For example, Anderson's
theorem for weak impurities only works for $s$-wave superfluids as we
study here.
 
\begin{figure}
\centerline{\includegraphics[width=3.3in,clip]{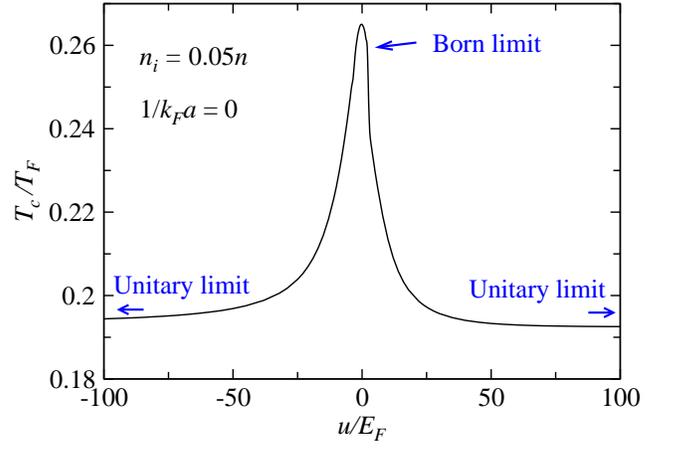}}
\caption{Superfluid transition temperature $T_c$ of a unitary Fermi
  gas as a function of impurity scattering strength $u$, from the Born
  limit through the unitary limit, with an impurity density
  $n_i=0.05n$. }
\label{fig:varu_on_Tc}
\vspace*{0.5cm}
\end{figure}

Next, we study the effects of impurity scattering strength on
$T_c$. Shown in Fig.~\ref{fig:varu_on_Tc} is $T_c$ as a function of
$u/E_F^{}$, from $-\infty$ to $+\infty$, for a unitary Fermi gas, with
a representative impurity density $n_i=0.05n$.  Here the impurity
strength $u$ spans from the Born limit to the unitary limit, for both
attractive and repulsive scatterers.  The $T_c$ value in the
$u\rightarrow 0$ limit is slightly higher than the clean system value,
$T_c^0/T_F^{} \approx 0.256$ \cite{ParticleHole}, due to the
subtraction of $\dbar{G}^R_{\omega=0}$ from $\dbar{G}^R_\omega$ in
Eq.~(\ref{eq:regularization}). In the Born regime, $T_c$ decreases
slowly with increasing $|u|$. Once $|u|$ increases further away from
the Born limit, $T_c$ decreases rapidly (as $\gamma \propto u^2$) at
first, then slows down and eventually approaches a constant in the
unitary scattering limit. Such asymptotic behavior is indeed
consistent with the expressions for $\Sigma_\omega$ and $\Gamma_0
(\omega)$ in Eqs.~(\ref{eq:renorm})-(\ref{eq:gamma0}), from which one
can readily show that in the large $|u|$ limit both $\Sigma_\omega$
and $\Gamma_0 (\omega)$ becomes essentially $u$ independent. The
suppression of superfluidity and $T_c$ by strong impurity scattering
is basically caused by two effects. On the one hand, strong impurity
scattering leads to a finite lifetime of fermionic quasiparticles, and
thus depletes DOS in the coherence peak and transfers spectral weight
to the impurity band and subgap states; such a spectral weight
relocation is detrimental to superfluidity. On the other hand, the
impurity scattering also causes a finite lifetime of fermion
pairs. While the former is dominant in the BCS regime, the latter
dominates the BEC side.

Note that the $T_c$ curve in Fig.~\ref{fig:varu_on_Tc} is almost
symmetric about $u=0$, consistent with Eqs.~(\ref{eq:ImpT}). This
should be contrasted to the case of $d$-wave superfluidity, where such
a symmetry is clearly absent due to vanishing
$\dbar{F}_\omega$ \cite{Chen-PRB}.

\subsection{Gaps and the superfluid density in the presence of impurities}
\label{subsec:superfluid_density}

\begin{figure}
\centerline{\includegraphics[width=3.3in,clip]{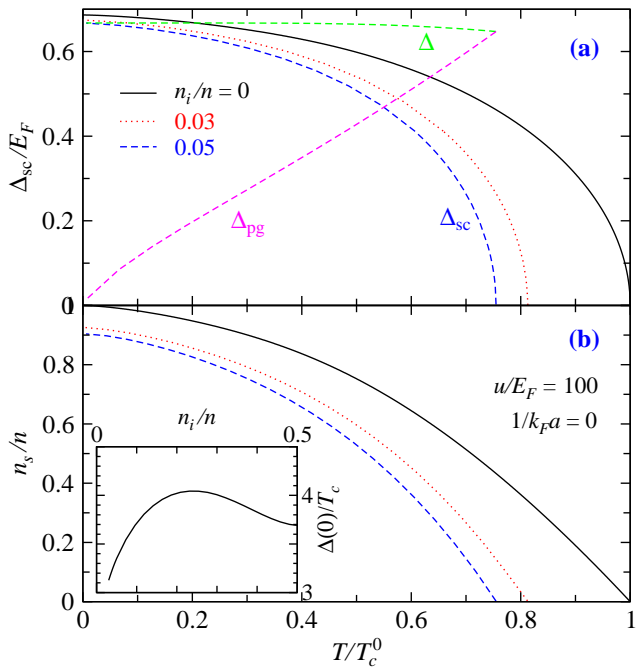}}
\caption{(a) Superfluid gaps and order parameter $\Delta_{sc}$ and (b)
  superfluid density $n_s/n$ in a unitary Fermi gas, as a function of
  $T/T^0_c$, for different densities of impurity, including the clean
  limit (black solid line) and $n_i/n=0.03$ (red dotted) and 0.05
  (dashed lines), with scattering strength $u/E_F^{}=100$. For clarity,
  we show the total gap $\Delta/E_F^{}$ and the pseudogap
  $\Delta_{pg}/E_F^{}$ only the $n_i/n=0.05$ case. Shown in the inset is
  a continuous evolution of the corresponding ratio $\Delta(0)/T_c$ vs
  $n_i/n$.  Here $T_c^0/T_F^{} = 0.256$ is the $T_c$ in the clean
  limit.}
\label{fig:gaps-Ns-vs-T}
\end{figure}

Now we investigate the transport properties of a Fermi gas in the
presence of impurities. The superfluid density can be derived using a
linear response theory. Following Ref. \cite{Chen-PRB}, for $s$-wave
pairing with $\varphi_\mathbf{k}=1$ in three dimensions, we obtain
\begin{eqnarray}
\frac{n_s}{m}\!\! &=&\!\! \frac{4}{3} \Delta_{sc}^2\! \sum_\mb{k}\!\! \int_{-\infty}^\infty\!\!
\frac{d\omega}{\pi}\; \mathrm{Im} \!\left(\!\tilde{F}^A(\omega,
    \mb{k})\!\right)^2\!\!
  (\vec{\nabla}\epsilon_\mb{k})^2  |\Gamma^R_0 (\omega)|^2 f(\omega), \nonumber \\ 
\label{eq:Ns}
\end{eqnarray}
where $\tilde{F}^A(\omega, \mathbf{k})=1/C^* \left( \omega, \mathbf{k}
 \right)$, different from $F(K)$ by a factor $\tilde{\Delta}_\mb{k}$, 
and $C\left( \omega, \mathbf{k}\right)$ is
given in Sec.~\ref{sec:IIB}. 

First we plot in Fig.~\ref{fig:gaps-Ns-vs-T}(a) the gaps and the order
parameter $\Delta_{sc}$ as a function of $T/T^0_c$ in a unitary Fermi
gas, where $T^0_c$ is the clean system superfluid transition
temperature. We choose strong impurities in the unitary limit, with
$u=100E_F^{}$, and calculate for three representative impurity densities
of $n_i/n = 0$ (clean, black solid curve), $0.03$ (red dotted curve),
$0.05$ (blue dashed curve), respectively.  For clarity, we show the
pseudogap $\Delta_{pg}$ and the total excitation gap $\Delta$ only for
the $n_i/n=0.05$ case. It is easy to conclude that unitary impurities
significantly suppresses both $T_c$ and the gaps, including the order
parameter. In addition, the suppression is more effective for $T_c$
than for the gaps. This can also be seen from the ratio
$\Delta(0)/T_c$ as a function of $n_i/n$, as shown in the inset; the
ratio initially increases rapidly with $n_i$, and drops slightly after
reaching a maximum. Such a non-monotonic and non-constant behavior
signals the breakdown of Anderson's theorem for strong impurities. We
note here that this ratio is rather different from the mean-field
value of 1.76, due to primarily the presence of a pseudogap at $T_c$
besides the impurity effects. 

\begin{figure}
\centerline{\includegraphics[width=3.in,clip]{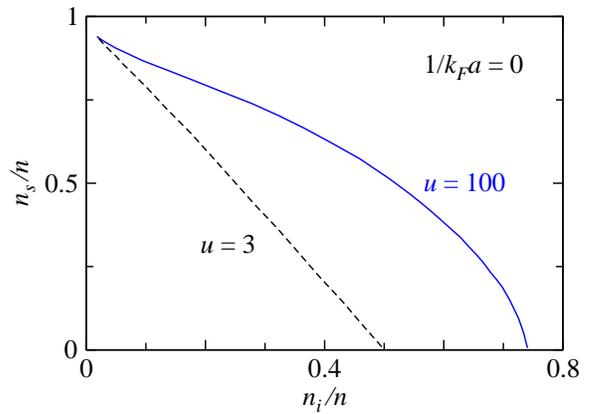}}
\caption{(a) Fractional superfluid density $n_s/n$ in a unitary Fermi
  gas at $T=0$ as a function of the relative impurity density $n_i/n$,
  for both unitary (blue solid) and Born (black dashed curves)
  scatterers. }
\label{fig:Ns-vs-Ni}
\end{figure}

With the calculated gap parameters, we show the corresponding
calculated superfluid density as a function of temperature for the
above the three impurity densities in Fig.~\ref{fig:gaps-Ns-vs-T}(b).
As with the order parameter, the superfluid density is suppressed
effectively by impurities in the unitary limit. While the reduction of
$n_s$ increases with $n_i$, nonlinearity is clearly present.  Detailed
study of $n_s$ versus $n_i$ at zero $T$ is shown in
Fig.~\ref{fig:Ns-vs-Ni}. Similar to the $d$-wave case on a lattice
\cite{Chen-PRB}, here we also find that for unitary scatterers
($u=100E_F^{}$), $n_s$ drops initially very fast with $n_i$ and then
slows down as $n_i$ increases further. In contrast, for Born
scatterers ($u=3E_F^{}$), $n_s$ decreases roughly linearly with $n_i$. It
should be noted that due to the large gap size at unitarity, it takes
a large $n_i/n$ in both cases to destroy the superfluid
completely. While the theory may break down at such a large impurity
density, it does indicate the robustness of an $s$-wave unitary Fermi
gas against impurity scattering, as compared to a weak coupling BCS
case (see e.g., Fig.~\ref{fig:CrossoverTc}) or the $d$-wave case shown in
Ref.~\cite{Chen-PRB}.

Finally, we note that it requires some effort to realize a homogeneous
Fermi gas in the presence of impurities. First, atomic Fermi gases are
always confined in a trapping potential. The impurities should be
confined within this trapping potential as well. To improve the
situation, one may use a combination of different lasers to create a
rather flat trap to make the system as close to homogeneous as
possible. Second, while the impurities may be realized by doping with
heavy atoms, a more elaborate treatment may need to consider the
finite mass of these impurity atoms. In this way, a different impurity
scattering Hamiltonian will have to be used. Third, an alternative to
realize impurities is to create (pseudo-)randomly distributed optical
speckles in the trap. In this case, these speckles may be regarded as
infinitely heavy impurities so that the scattering of atoms is
elastic, as assumed in our theory. Nevertheless, the present theory
can be regarded as a first step toward a more realistic treatment of a
real Fermi gas system with random impurities. Spin flip (i.e.,
``magnetic'') impurity scattering will also be considered in future
works. Furthermore, we shall also include the particle-hole channel
contributions \cite{ParticleHole} as well.

\section{SUMMARY}
\label{sec:IV}

In summary, we have studied the effects of impurities on the $s$-wave
BCS-BEC crossover in ultracold atomic Fermi gases, including the
impurity effects on frequency and gap renormalizations, fermion
density of states, superfluid $T_c$, as well as finite temperature
gaps and superfluid density. Our results reveal that while the system
is less sensitive to impurities in the Born limit, strong impurities
in the unitary scattering regime cause a much stronger renormalization
for both the frequency and the gaps throughout the entire BCS-BEC
crossover, leading to a finite lifetime of Bogoliubov quasiparticles
and fermion pairs, and hence a significant suppression of the
superfluid $T_c$ and superfluid density. The Anderson's theorem breaks
down except for weak impurities in the BCS regime. Indeed, in the weak
coupling BCS regime, where the gap is small, strong impurities at
moderately high densities may readily destroy the superfluidity and
suppress $T_c$ down to zero, leading to an effective power law
dependence on the pairing strength.  Such a BCS-BEC crossover
phenomenon in the presence of impurities may be realized
experimentally by introducing atoms of foreign elements or using
optical speckles, with Feshbach resonance in Fermi gases of $^6$Li or
$^{40}$K.

\acknowledgments

This work is supported by NSF of China (Grant No. 11274267), the
National Basic Research Program of China (Grants No. 2011CB921303 and
No. 2012CB927404), and NSF of Zhejiang Province of China (Grant No.
LZ13A040001).

\bibliographystyle{apsrev}

\end{document}